\documentclass[usenatbib]{mn2e}
\usepackage{aas_macros,epsfig,amsmath}

\newcommand{\lesssim}{\tilde{<}}
\newcommand{\gtrsim}{\tilde{>}}

\newcommand{\mpc}{\, {\rm Mpc}}

\newcommand{\hmpc}{\, h^{-1} \mpc}
\newcommand{\ihmpc}{\, h\, {\rm Mpc}^{-1}}

\newcommand{\PNL}{P_{\rm NL}(k,z)}

\newcommand{\weffect}{P_{\rm NL}(w)/P_{\rm NL}(w=-1)}

\begin{document}
\topmargin-1cm

\title{ Dependence of the non-linear mass power spectrum on
the equation of state of dark energy }

\author[Patrick McDonald, Hy Trac, and Carlo Contaldi]
{Patrick McDonald$^1$\thanks{Electronic address: 
    {\tt pmcdonal@cita.utoronto.ca} 
},
Hy Trac$^{2,3}$ and Carlo Contaldi$^{1,4}$
\\$^1$ Canadian Institute for Theoretical
Astrophysics, University of Toronto, Toronto, ON M5S 3H8, Canada
\\$^2$ Department of Physics, Jadwin Hall, Princeton University, Princeton,
NJ 08544, USA
\\$^3$ Princeton University Observatory, Princeton University, Princeton NJ
08544, USA
\\$^4$ Imperial College, South Kensington Campus, London, SW7 2BW, U.K.
}

\maketitle

\begin{abstract}

We present N-body simulation calculations of the dependence of the power 
spectrum of non-linear cosmological mass density fluctuations on the equation 
of state of the dark energy, $w=p/\rho$.  At fixed linear theory power, 
increasing $w$ leads to an increase in non-linear power, with the effect 
increasing with $k$.  By $k=10~\ihmpc$, a model with $w=-0.75$ has 
$\sim 12$\% more power than a standard cosmological constant model ($w=-1$), 
while a model with $w=-0.5$ has $\sim 33$\% extra power (at $z=0$).  The size 
of the effect increases with increasing dark energy fraction, and to a lesser 
extent increasing power spectrum normalization, but is insensitive to 
the power spectrum shape
(the numbers above are for $\Omega_m=0.281$ and $\sigma_8=0.897$).  
A code quantifying the 
non-linear effect of varying $w$, as a function of $k$, $z$, and other 
cosmological parameters, which should be accurate to a few percent for 
$k\lesssim 10~\ihmpc$ for models that fit the current observations, is 
available at http://www.cita.utoronto.ca/$\sim$pmcdonal/code.html.  This 
paper also serves as an example of a detailed exploration of the numerical 
convergence properties of ratios of power spectra for different models, which 
can be useful because some kinds of numerical error cancel in a ratio.  When 
precision calculations based on numerical simulations are needed for many 
different models, efficiency may be gained by breaking the problem into a 
calculation of the absolute prediction at a central point, and calculations of 
the relative change in the prediction with model parameters.  

\end{abstract}


\section{Introduction}

Dark energy is the most important focus of study in 
cosmology today.  Its basic existence is well established,
both by observations of Type Ia supernovae 
\citep{2003ApJ...598..102K,2004ApJ...607..665R}, 
and by combinations of other observables \citep{2005PhRvD..71j3515S}.
The focus now is on probing the properties of the dark energy
(or whatever new physics causes the  
Universe to look like it contains dark energy).
A simple first parameterization of its
properties is to specify the ratio of pressure to density 
(equation of state) $w=p/\rho$, with $w=-1$ for a cosmological 
constant.

The large amount of observational effort focused on measuring $w$ to 
high precision must be matched by sufficiently accurate predictions
of the dependence of observables on $w$.  Theory 
calculations will need to be done more carefully than they have been 
in the past.  In particular, probes of $w$ based on cosmological  
structure (e.g., weak lensing, galaxy clusters, and even 
large-scale galaxy clustering where the linear power assumption is no
longer exclusively relied on 
\citep{2004PhRvD..69j3501T,2004astro.ph..8003A,2005astro.ph..1171E})
will generally require non-linear numerical simulations
as the fundamental method of calculating theory predictions.  
This paper tackles a small part of the problem:  computing 
the dependence of the non-linear mass power spectrum, $\PNL$,
on $w$.  

The most direct use of calculations of the non-linear power spectrum is for
weak gravitational lensing (cosmic shear) studies (e.g.,
\cite{2003astro.ph..6033B,2004astro.ph.11673S,2005APh....23..369H,
2005astro.ph..2243J,2005astro.ph..3644K}).
Direct use of the mass power for weak lensing (as opposed to
ray tracing) is of course only as good as the Born and Limber approximations.
\cite{2003ApJ...592..699V} found the Born approximation agreed well with
ray tracing, although this is less clear in \cite{2004APh....22...19W}.
In a pilot project, \cite{2004APh....22...19W} performed a small grid of 
simulations of weak lensing with full ray tracing, including
$w=-0.8$, for parameter combinations designed to leave the CMB
fluctuations invariant.  When performing a full grid of models for
precision parameter fitting, this CMB-guided method is probably the best way 
to go.  Here we concentrate
on isolating the effect of $w$ by running simulations where only 
$w$ and possibly one other parameter is varied at a time 
(although the fitting
formula we present does include joint variations of $w$, $\sigma_8$,
and $\Omega_m$).
Even if ray tracing is ultimately required for high precision, it should 
still be useful to have an accurate mass power spectrum
calculation, so we leave ray tracing for future work.

The fitting formulas most commonly used to predict the non-linear
power spectrum given a set of cosmological parameters were not 
calibrated with $w\neq -1$ simulations 
\citep{1996MNRAS.280L..19P,2003MNRAS.341.1311S},
although \cite{2001PhRvD..64h3501B} present an 
untested prescription for using \cite{1996MNRAS.280L..19P}.
Often these formulas are used for $w\neq -1$
by making the untested assumption that models with equal linear
theory power and other parameters (most importantly $\Omega_m$) at the 
redshift of interest will have equal non-linear power, independent of
$w$. 
\cite{1999ApJ...521L...1M} did simulate the mass power spectrum for $w=-2/3$,
$-1/2$, and $-1/3$, and combined these with the $\Lambda$CDM simulations
in \cite{1998ApJ...508L...5M} to produce a fitting formula.  There is no
obvious reason not to trust this formula to the 10\% level of accuracy
advertised; however, the grid of simulations used to calibrate it was 
sparse (e.g., the $w>-1$ simulations all had $\Omega_m\geq 0.4$), 
and numerical convergence was not rigorously demonstrated.  We will
find that this formula does not work well in the region of parameter
space where we find ourselves. 
\cite{2003ApJ...599...31K} simulated
models with different values of $w$, but did not present the power 
spectrum in detail.  The one case they did show, a 
\cite{1988PhRvD..37.3406R} model with time varying $w\sim -0.5$, 
appears to be roughly consistent with our results. 

We do not aim in this paper to replace the existing fitting formulas 
for the non-linear power  --
only to provide an accurate correction for $w\neq -1$.  
We generally consider flat models with parameters $\sigma_8$ (the 
rms linear 
mass density fluctuations in $8\hmpc$ radius spheres at $z=0$), 
$\Omega_m$, $\Omega_b$ (the mass and baryon densities relative to the 
critical density), $h$ (the Hubble parameter), $n$ (the logarithmic 
slope of the primordial power spectrum), and $w$. 
The dark energy equation of state parameter
$w$ does not affect the shape of the transfer function on non-linear 
scales (e.g., $<0.5$\% difference at $k>0.007~\ihmpc$ between the 
transfer function for $w=-1$ and $w=-1/2$, from CMBFAST 
\citep{1996ApJ...469..437S}, for a typical model normalized at high-$k$), 
so the $z=0$ linear theory power is not changed at all by changing $w$, when 
our other parameters are fixed.  There can be, however, a change in the
non-linear power with $w$, because the growth history changes -- increasing 
$w$ means the Universe had relatively more linear theory power at earlier 
times.  This effect is not included in an estimate of the non-linear power 
made using something like the \cite{2003MNRAS.341.1311S} fitting 
function.

We will frequently refer to the change in observables with $w$ at fixed $z=0$ 
values of other parameters as ``the effect of $w$.'' We are not implying that 
there is anything uniquely correct about this.  
From some points of view it would make more sense
to fix the other parameters at early times, before the dark
energy has become 
significant.  We focus on the former definition of the effect of $w$ 
simply because it is generally not included in
weak lensing calculations \citep{2005APh....23..369H,2004astro.ph.11673S,
2005astro.ph..2243J,2005astro.ph..3644K}, and
can not be reproduced in any obvious way using the existing codes.
While it is mostly a matter of arbitrary definition, there is a real 
sense in which our choice is ``the'' non-linear effect of $w$ at $z=0$:  
our effect goes to zero in the limit of small perturbations.

The range of $k$ and $z$ we consider, and the accuracy goal, are guided by
the weak lensing application 
($k\equiv 2 \pi/\lambda$, where $\lambda$ is the 
wavelength of a Fourier mode).
We limit ourselves to the range $0<z<1.5$, because this range is most directly
sensitive to the presence of dark energy and most 
relevant to galaxy weak lensing surveys, and because the power at 
increasing redshift should probably be simulated using decreasing box size,
since limited particle density becomes an increasingly serious problem
(the true small-scale power is smaller relative to the spurious 
particle discreteness-related power)
while limited box size becomes less problematic (smaller scale modes are
still linear).
\cite{2005APh....23..369H} investigated the requirements on the 
$\PNL$ calculation for future large weak lensing
surveys, finding that 1-2\% accuracy should be sufficient, 
or 0.5\% in the worst possible case.  They found that the 
relevant scales are $0.1 \lesssim k \lesssim 10 \ihmpc$.
\cite{2004ApJ...616L..75Z} studied the effect of hot baryons on the
weak-lensing shear power spectrum in halo models.
They found an effect of roughly 5-10\% on the power at $k=10~\ihmpc$
(reading from their Figure 1),
so trying to achieve better than a few percent precision at this $k$
using simulations without gas dynamics is probably pointless (it is
always good to aim for errors somewhat smaller than the other known
sources).  \cite{2004PhDT.......411Z} found roughly similar results 
in SPH simulations (reading from Figure 5.7), and 
\cite{2004APh....22..211W} found a
comparable scale for the effect of baryonic cooling.
We will not actually achieve 1-2\% level accuracy in the dependence 
of power on $w$, but we think we show a clear path to it.  The 
accuracy we do achieve is substantially better than anything else 
available, so it should be useful until a larger project can 
improve it.
 
Throughout the paper, we use the trick of canceling numerical errors by taking 
ratios of power spectra for different models, rather than looking at the 
absolute power in each model directly.  This should
in no way be considered a swindle or an added approximation.  
It is completely reasonable and
expected that many types of error are insensitive to the model, so
that ratios (or differences) between the models can genuinely be
computed more accurately than either model individually.  One example
of this is the effective smoothing involved in mapping particles to a 
grid so that you can FFT the periodic density field
for the purpose of
measuring the power.  This suppresses the
power substantially, but by precisely the same factor in all models.
This factor cancels exactly when we take a ratio, up to some high $k$ where
the power may be corrupted by aliasing, or the suppression is simply too
large to invert accurately. 
The key to believing these results is that the convergence
of the ratios of power with numerical parameters of the simulations
must be tested carefully in the same way a direct measurement of the
power would be.  For example, taking ratios would not cancel an 
additive Poisson noise term due to the limited number of particles,
but this would become completely obvious in a convergence test where
the number of particles is varied.  

We use a particle-multi-mesh (PMM) code, based on an improved
particle-mesh (PM) algorithm, for our grid of N-body simulations.  In
principle, PM codes can achieve high spatial resolution but at a great
cost in memory and to a lesser extent in work.  In practice, they are
normally limited to a mean interparticle spacing-to-mesh cell spacing
ratio of 2:1, where the storage requirements for particles and grids
are approximately balanced.  PMM utilizes a domain-decomposed,
FFT-based gravity solver 
\citep{2004NewA....9..443T,2005NewA...10..393M} to
achieve higher spatial resolution while maintaining memory costs.  We
will find that a ratio of 4:1 between the particle grid spacing and
mesh grid spacing is roughly optimal, in the sense 
that at a given $k$ 
(the most relevant $k$ where the errors are a few percent) the error 
from limited
particle density and limited force resolution are similar, for ratios
of power spectra from simulations with different $w$. 
We note that \cite{2004astro.ph.11795H} compared several N-body codes
(not including ours),
looking at absolute power, and found good general agreement.

It can be useful to consider numerical errors rather carefully,
focusing on the initial small breakdown of accuracy, because, 
for example, a factor of two change in resolved scale or required box size
can easily change the require computer time or memory by a factor of 
at least eight.  
At least five things must be investigated in every simulation 
program:  
Sensitivity to starting redshift, box size, mass resolution (i.e.,
number of particles), force resolution, and time step size.
We also test our method of power spectrum calculation.
Even though not everything we find will be completely generalizable to other
types of simulations,
we think it is useful to lay out an attempt to push them all to a 
percent level of control.  

Throughout this paper we use the philosophy of suppressing our
urge to be general (e.g., increase the redshift range and parameter space
coverage and study weak lensing directly) in favor of patiently 
investigating a restricted problem.
I.e., rather than trying to entirely solve the problem of calculating 
the power spectrum over all parameter space, which at this point would
require approximations and cutting corners,
we select a small subproblem and explore
it carefully, in hopes of learning better how to do the full problem
accurately and efficiently.

Recently \cite{2005MNRAS.357..387K} performed a set of simulations 
of models with different $w$ values, although they focused on dark 
matter halo density profiles.  We mention them because 
they also represent a good source
for some pedagogical figures, e.g., of the linear transfer function 
and growth factor as a function of $w$.

The rest of the paper is laid out as follows:
First, in \S 2, we qualitatively demonstrate the effect of changing 
the equation of 
state of the dark energy on the non-linear mass power spectrum.  
Then, in \S 3, we describe detailed tests of our simulations (this 
section is not intended for the casual reader). 
Finally, in \S 4, we describe the code we provide to quantify the 
results.

We often describe simulations using
the notation $(L,~P,~M)$, where $L$ is the box size in comoving
$\hmpc$, $P^3$ is the number of particles, and $M^3$ is the number of mesh 
cells.

\section{The effect of changing $w$ \label{secw}}

\begin{figure}
\centerline{\psfig{file=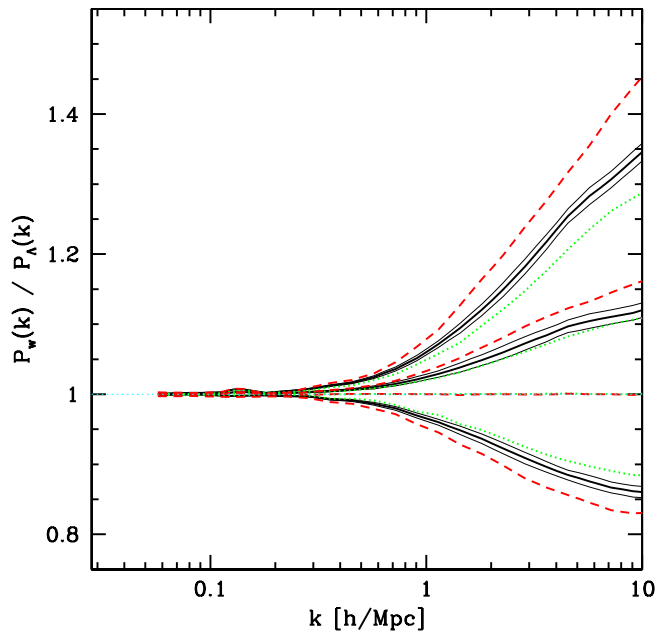,width=3.5in}}
\caption{
Fractional effect of $w$ on the non-linear mass power spectrum,
at fixed linear theory power.
The thick lines of a given color/type
show, from top to bottom, $w=-0.5,-0.75, -1.5$, all relative to $w=-1$.  
The black, solid line is for $\Omega_m=0.281$, with red, dashed (green, dotted)
showing $\Omega_m=0.211$ (0.351).  Thin black lines show rms 
statistical error bands (in Figures \ref{basiceffect}-\ref{z1.5eff}, the 
statistical errors are all similar to these examples).
}
\label{basiceffect}
\end{figure}

Figure \ref{basiceffect} shows the basic effect of $w$ on the non-linear
power, at $z=0$.   
To minimize the statistical fluctuations, we have used the same initial 
conditions for simulations with different $w$, except that we adjust the
initial amplitude of the perturbations by the factor necessary to produce 
an identical linear theory density field 
at $z=0$ (and as a result identical $\sigma_8$).
All of the figures in this section show results from (110,192,768) 
simulations.
In the next section we investigate the accuracy of the results in detail, 
finding statistical and numerical convergence to a few percent.
Inevitably, the effect of $w$ increases with decreasing $\Omega_m$, i.e., 
increasing 
dark energy fraction, as shown in the figure.
To be clear:  in our many 
figures like this, we are plotting the ratio of power in a model
with $w\neq -1$ to a model with $w=-1$, with all the other parameters
the same at $z=0$ in both.  For example, when we show the result for 
a different 
value of $\Omega_m$, $\Omega_m$ has been changed for both the model in the 
numerator and the model in the  denominator 
For notational compactness,
we will sometimes refer to this type of ratio as 
$f_{w}(k)\equiv P_{w}(k)/P_{w=-1}(k)$.  

To put the 12\% (33\%) effect of $w=-0.75$ ($w=-0.5$) at $k=10\ihmpc$ 
in perspective, we note that the ratio of linear growth of power 
from early times to $z=0$ between the $w=-1$
and $w=-0.75$ (-0.5) models is $D^2(w)/D^2(w=-1)=0.83$ (0.57).  Note that
weak lensing measurements tend to be most sensitive to somewhat smaller $k$
\citep{2005APh....23..369H}.  On the other hand, 
\cite{2005astro.ph..4557H} recently
showed that a modification of the power similar in form to ours has 
a substantial effect on the convergence power over a wide range in 
$\ell$.  
Considering the non-trivial $k$
and $z$ dependence, a full parameter forecast calculation 
(e.g., \cite{2004astro.ph.11673S}) will be 
needed to see if this effect can change the projections for future weak 
lensing measurements of $w$
significantly.

We have compared this
result to the formula of \cite{1999ApJ...521L...1M}, 
and the agreement is not good.  
The \cite{1999ApJ...521L...1M} formula
predicts a much larger effect of $w$, with $f_{w=-0.5}(k)$ passing 1.5 at
$k=0.4\ihmpc$, before plateauing at a value of $\sim 2$.   
This disagreement is not too surprising since 
\cite{1999ApJ...521L...1M} did not use
simulations in which $w$ was varied at fixed values of the other parameters.
Without a very comprehensive grid of simulations, the simple dependence of 
the non-linear power on the linear power (and $\Omega_m$) could easily 
be mixed with the type of $w$ dependence that we are studying here.
We remind the reader that no rigorous analytic prediction
for the non-linear power spectrum exists.
The \cite{1999ApJ...521L...1M} or \cite{2003MNRAS.341.1311S} formulas
are not derived from first principles -- they are physically motivated 
fitting functions that are used
to interpolate/extrapolate simulation results.  Even if their basic motivation
is perfectly valid, they contain completely free parameters that are only
determinable through fits to simulations, so their predictions are only as
good as the simulations used to calibrate them.  \cite{1999ApJ...521L...1M} 
used only five 
models, with ($w$, $\sigma_8$, $\Omega_m$, $h$) = 
(-1, 1.29, 0.3, 0.75),
(-1, 1.53, 0.5, 0.7), (-2/3, ?, 0.4, 0.65), (-1/2, ?, 0.4, 0.65), and
(-1/3, ?, 0.45, 0.65).  The value of $\sigma_8$ was not given for the $w>-1$ 
simulations, but
they were COBE normalized \citep{1997ApJ...480....6B}, 
meaning they all have different $\sigma_8$.
One might hope that this was enough to calibrate
the fitting formula everywhere, but this is not at all obvious.
It would be very difficult using this set of simulations to disentangle the
effects of the different parameters well enough to accurately 
{\it extrapolate}
to a point in parameter space not bounded by the set.  In particular,
a parameter that has a relatively small direct effect will be most difficult
to control.  

\begin{figure}
\centerline{\psfig{file=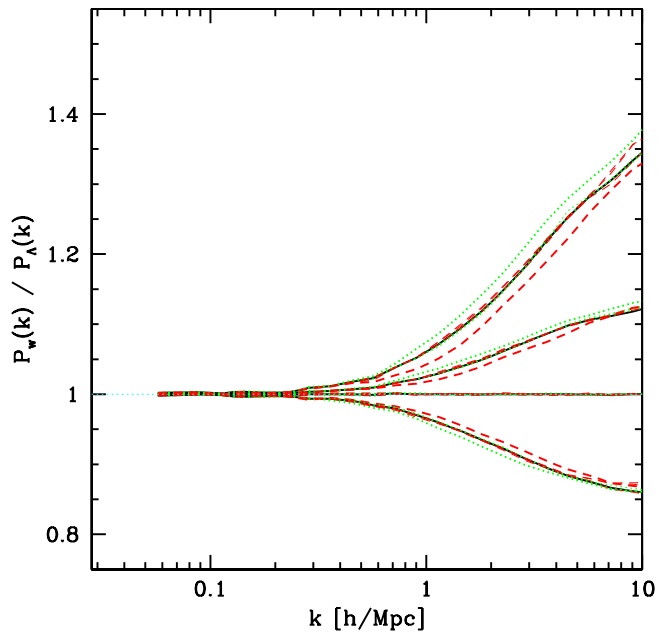,width=3.5in}}
\caption{
Similar to Figure \ref{basiceffect}, except
thick red, dashed (green, dotted) lines show $\sigma_8=0.800$ (0.994).   
(our central model has $\sigma_8=0.897$).
Thin lines for variations $\Omega_b=0.0462 \pm 0.0052$, $h=0.710\pm 0.066$, 
and $n=0.980\pm 0.065$ are also plotted, to show that they are usually 
indistinguishable from the central model.
}
\label{sig8effect}
\end{figure}

The normalization of the power spectrum, $\sigma_8$, also affects the result 
in a small, qualitatively reasonable way, with increasing $w$ dependence for 
increasing $\sigma_8$,
as shown in Figure \ref{sig8effect}
(the non-linear effect must disappear as $\sigma_8\rightarrow 0$). 
Figure \ref{sig8effect} also shows that the changes in the shape of
the power spectrum that arise from changes in $\Omega_b$, $h$, 
and $n$ do not
change the $w$ dependence significantly.  

\begin{figure}
\centerline{\psfig{file=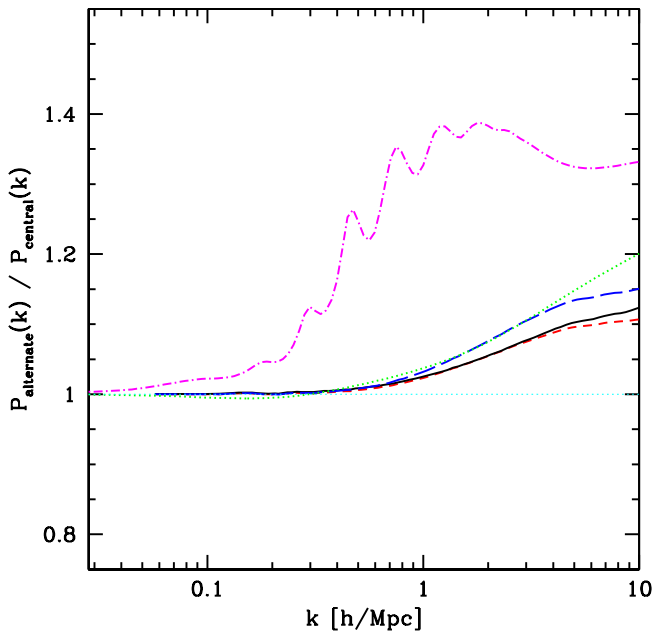,width=3.5in}}
\caption{
Comparison of the effect of changing $w$ to the effect of changing $\Omega_m$, 
at fixed $z=0$ linear theory power.
Black, solid line:  our standard ratio of simulated $P_{\rm NL}(k)$ 
with $w=-0.75$ to $w=-1$.
Blue, long-dashed line:  ratio of power with $\Omega_m=0.192$ to our standard
$\Omega_m=0.281$ (both with $w=-1$).  
These two alternative models have identical linear 
theory power at both $z=24$ and $z=0$.  
Red, short-dashed line:  ratio of power with $\Omega_m=0.213$ to standard
(this case has linear power equal to the $w=-0.75$ case at $z=1.9$).  
Green, dotted line:  as blue, but based on 
Smith et al. (2003) 
(for both numerator and denominator).  
Magenta, dot-dashed:  as blue but based on 
Ma et al. (1999). 
}
\label{omegasmith}
\end{figure}

To explore the origin of the $w$ effect, in Figure \ref{omegasmith} 
we show a similar calculation of the effect of changing 
$\Omega_m\equiv 1-\Omega_\Lambda$ 
while holding the $z=0$ linear power fixed (along with all the other
parameters, including $w=-1$).
Reducing $\Omega_m$ to 0.192 (from 0.281) produces a model with the
same power at $z=24$ (our starting redshift)
as the model with $w=-0.75$ and $\Omega_m=0.281$.  
We see that the
non-linear power in these two models is quite similar.  When we
match the linear power at $z=1.9$, using $\Omega_m=0.213$, the 
non-linear power is even more similar.  It seems likely that  
Figure \ref{omegasmith}, and the effect of $w$ in general, 
could be interpreted in a halo model
by accounting for the difference in formation redshift, and thus
density profile, of the typical halos dominating the power
on the scale of interest 
\citep{2005MNRAS.357..387K,2003MNRAS.340.1199H,2005NewAR..49..199B,
2004A&A...416..853D,2003A&A...400...19B,2002A&A...396...21B}.

The code of \cite{2003MNRAS.341.1311S} should be able to
reproduce this $\Omega_m$ dependence.  As we see in Figure \ref{omegasmith},
the agreement is excellent (we can not say for sure that our 
simulations are
accurate enough to believe the $\sim 5$\% disagreement at $k=10\ihmpc$, 
although they probably are).
Note that to make this comparison we have modified 
the \cite{2003MNRAS.341.1311S} code to accept 
an input linear theory
power spectrum in place of its usual calculation based on 
\cite{1984ApJ...285L..45B}.
The formula of \cite{1998ApJ...508L...5M,1999ApJ...521L...1M} 
does poorly in this test, probably
because its $\Omega_m$ dependence was essentially calibrated by two 
simulations with $\Omega_m=0.3$, $\sigma_8=1.29$, and 
$\Omega_m=0.5$, $\sigma_8=1.53$, i.e., far from the combination of 
parameters we are testing.

\begin{figure}
\centerline{\psfig{file=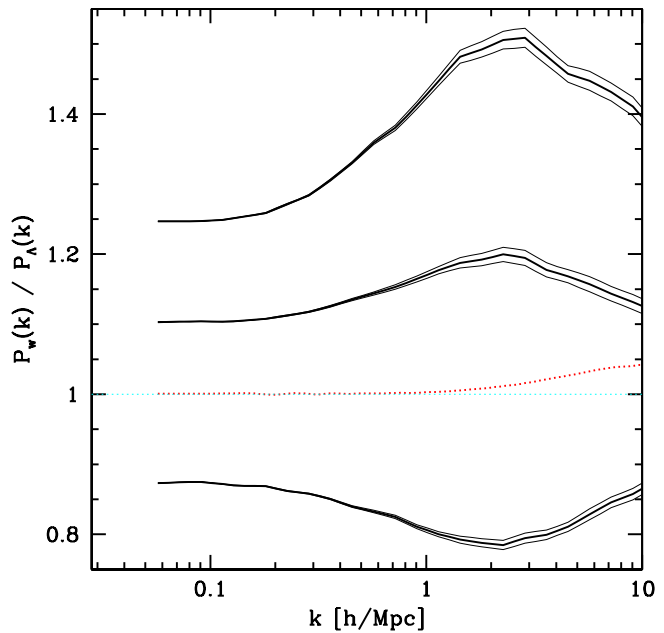,width=3.5in}}
\caption{
Non-linear power for varying $w$ at $z=1.5$. Thick black (solid) lines are for 
models with identical linear power and $\Omega_m$ at $z=0$ 
(from top to bottom $w=-0.5$, -0.75, and -1.5, all relative to $w=-1$).  
Thin black lines show the rms statistical error bands.
Note that much of the change here is accounted for by simple differences 
in linear growth.
The red (dotted) line shows $w=-0.5$ relative to $w=-1$ when the two models
have been constructed to have identical linear power and $\Omega_m(z)$ at 
$z=1.5$.  
}
\label{z1.5eff}
\end{figure}

Finally, in Figure \ref{z1.5eff} we show the effect of $w$ on the 
non-linear power
at $z=1.5$, at fixed values of our other parameters.
Note that in Figure \ref{z1.5eff} we are comparing models 
with different
linear theory power at the observed redshift, and different 
$\Omega_m(z)$.  
The difference between $P_{\rm NL}(z=1.5,~w=-0.5)$ and
$P_{\rm NL}(z=1.5,~w=-1)$ at fixed $z=1.5$ linear theory power and
$\Omega_m(z=1.5)$ is small, so it is not very interesting to plot this, 
but we show one case, comparing $w=-1.0$ with our usual linear theory
power spectrum at $z=0$, but $\Omega_m=0.150$, to $w=-0.5$ with 
$\sigma_8$ increased to 1.004, and $\Omega_m=0.411$ (these two models have 
exactly matching linear power and $\Omega_m(z)$ at $z=1.5$). 

\section{Numerical details \label{secdetails}}

In this section we investigate the dependence of our calculations on 
the numerical parameters of the simulations.   Beyond testing the
specific results we present,
we hope to contribute to the collective wisdom of the
research community about
how to do precision cosmology based on numerical simulations by 
exploring the idea of looking at the ratio of power in different
models in simulations using identical random
numbers for the initial conditions.

For our grid of N-body simulations, we used the PMM code, an improved
version of the PM algorithm.  It is based on a two-level mesh Poisson
solver where the gravitational forces are separated into long-range and
short-range components, as described in detail in  
\cite{2004NewA....9..443T} and \cite{2005NewA...10..393M}.  
The long-range force is computed on the
root-level, global mesh, much like in a PM code.  To achieve higher
spatial resolution, the domain is decomposed into cubical regions and
the short-range force is computed on a refinement-level, local mesh.
In the current version, PMM can achieve a spatial resolution of 4 times
better than a standard PM code at the same cost in memory.

Simulations with $P=192$ and $M=768$ fit 
conveniently into a single node on the 528-CPU Beowulf cluster at CITA,
and run in less than a day.  The importance when doing this kind of study
of complete freedom to run an arbitrarily large number of exploratory 
simulations,
with relatively quick turn-around, can not be underestimated, so we focus
on this configuration.  Future precision simulation grids will of course use
larger simulations.  
Our initial guess was that (220, 192, 768) simulations would be most useful,
so this section will focus first on the convergence properties of these,
including comparisons of (110, 96, 384) to
(110, 192, 384), to test the effect of finite particle density at the same
force resolution as the (220, 192, 768) simulations, and comparisons of
(110, 96, 384) to (110, 96, 768) to test the effect of force resolution.
Ultimately, we will conclude that we were overly worried about limited box
size and insufficiently worried about limited resolution, so subject to the
constraint $P=192$, $M=768$, somewhat smaller box size is optimal
(the code we release is based on (110, 192, 768) simulations).
 
Throughout this section our plots use a standard vertical axis range 
$0.94-1.06$, to allow easier comparison of the size of different errors.

\subsection{Initial conditions}

Our transfer functions are computed using the ``lingers'' 
code associated with 
grafic2-1.01 \citep{2001ApJS..137....1B}.
GRAFIC2 is then used to generate the initial conditions. 
We turn off
the GRAFIC2 Hanning window, which isotropizes the small-scale structure
at the expense of suppressing the small-scale power. 

To determine the linear growth factor (used to convert initial
conditions generated for $w=-1$ models into $w\neq -1$ models) 
we numerically solve
\begin{equation}
D^{\prime\prime} +\frac{3}{2}\left[1-\frac{w(a)}{1+X(a)}\right]\frac{D^\prime}{a}-
\frac{3}{2}\frac{X(a)}{1+X(a)}\frac{D}{a^2}=0
\end{equation}
\citep{2003MNRAS.346..573L}, with 
\begin{equation}
X(a)=\frac{\Omega_m}{1-\Omega_m}e^{-3\int_a^1 d\ln a^\prime w(a^\prime)}~.
\end{equation}

When we compare simulations with the same box size but different particle 
density, the initial conditions of the box with fewer particles are 
set by a sharp $k$-space filter applied to the initial conditions of the
higher resolution box.  This is equivalent to simply regenerating the 
initial conditions with the same random numbers for the large-scale modes
(for our method of generating initial conditions),
but not equivalent to re-binning the density and momentum fields in real space 
(the latter method introduces a suppression of high-$k$ power that increases 
the level of disagreement between the results for the two particle densities). 

\subsection{Starting redshift}

\begin{figure}
\centerline{\psfig{file=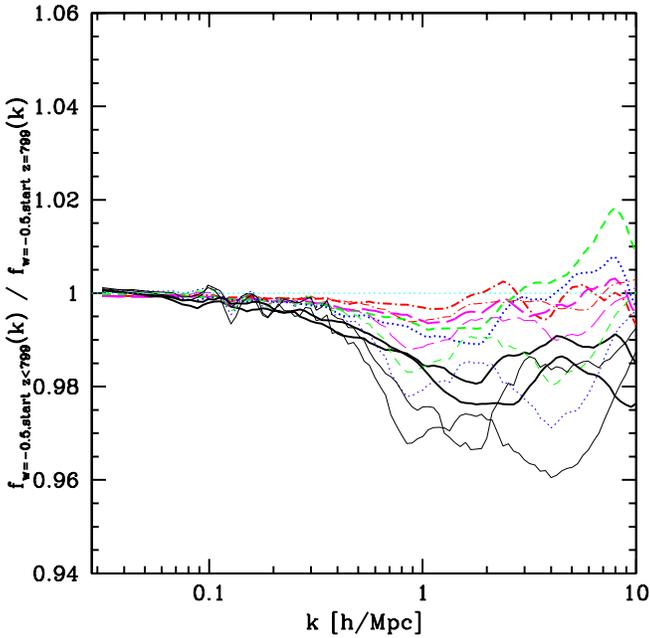,width=3.5in}}
\caption{
Change in the effect of $w$, $f_{w=-0.5}(k)\equiv P_{w=-0.5}(k)/P_{w=-1}$, 
with
starting redshift in the simulations.  Thick (thin) lines are power at 
$z=0$ ($z=1.5$). 
The denominator is for starting $z_{\rm i}=799$, with 
black/solid showing the
difference for $z_{\rm i}=24$, blue/dotted $z_{\rm i}=49$, green/short-dashed
$z_{\rm i}=99$, magenta/long-dashed $z_{\rm i}=199$, and red/dot-dashed
$z_{\rm i}=399$.
The two black lines of each type show different realizations of the initial
conditions.
}
\label{wzstart}
\end{figure}

It is important for precision calculations to test the affect of changing
the starting redshift in the simulations.  This can identify, for example,
transients due to the imperfection of the \cite{1970A&A.....5...84Z} 
approximation \citep{1998MNRAS.299.1097S}.
Figure \ref{wzstart} shows the change in the effect
of $w$ as we increase the starting redshift, $z_{\rm i}$, from our standard
$z_{\rm i}=24$.
Note that it is not automatically the case that higher $z_{\rm i}$ is better,
because numerical errors (most obviously suppression of power by limited force
resolution) have more time to accumulate in that case.  The agreement is
good but not perfect, with errors as large as 2\% at $z=0$ and $\sim 3$\% 
at $z=1.5$.  Subsequent to our runs for this paper, we found a small problem
in the PMM force kernel which leads to this disagreement -- it is not 
serious so we leave the more accurate calculation for the future. 

\subsection{Mass resolution}

\begin{figure}
\centerline{\psfig{file=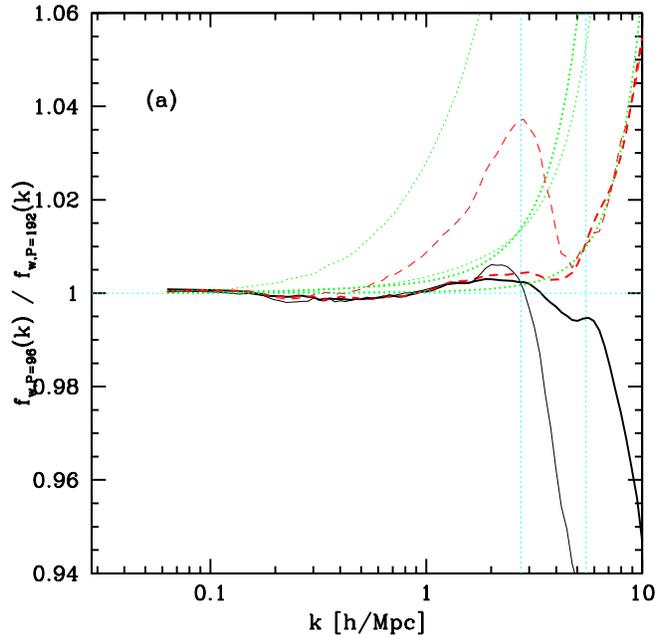,width=3.5in}}
\caption{
Change in the effect of $w$, $f_{w=-0.5}(k)\equiv P_{w=-0.5}(k)/P_{w=-1}$, 
with
number of particles.  Thick (thin) lines are power at 
$z=0$ ($z=1.5$), with $L=110\hmpc$ (a) or $L=55\hmpc$ (b)  
and force mesh $M=384$. 
The numerator is simulations with $(P=96)^3$ particles,
the denominator is $P=192$.
Red (dashed) lines include subtraction of Poisson noise 
$P_{\rm noise}(k)=(L/P)^3$,
black (solid) do not.
For reference, the green (dotted) lines 
show $1+P_{\rm noise}(k)/P_{\rm measured}(k)$, where
$P_{\rm measured}(k)$
is the measured power for the $w=-1$ model (after deconvolution of the 
power spectrum measurement
mesh, but without noise subtraction).  The two green lines of each thickness 
represent the two particle densities.
The vertical cyan (dotted) lines mark $k=(96,192)\pi/110 \hmpc$.
}
\label{particlenumber}
\end{figure}

Figure \ref{particlenumber}(a) shows the effect of varying the number of
particles, for a fixed force resolution, by comparing the average of 
two (110, 96, 384)
simulations to the average of two (110, 192, 384) simulations 
(the pairs have different random initial conditions).
\begin{figure}
\centerline{\psfig{file=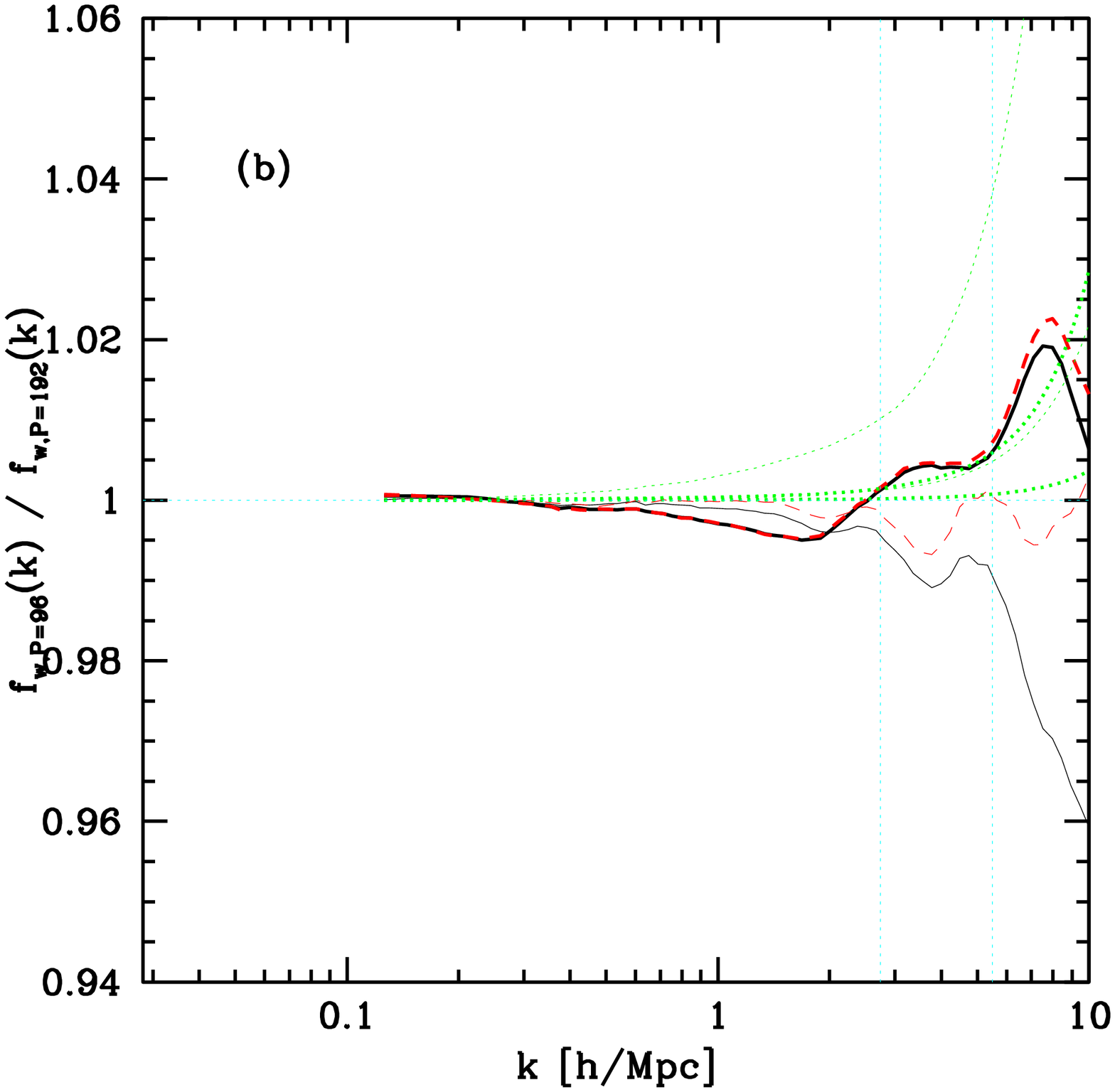,width=3.5in}}
\end{figure}
Note that this is a test of the effect of adding high $k$ power in 
the initial conditions in addition to the effect of simply 
subdividing the mass.
We have done the same 
comparison with a factor of two better force resolution and the results
are similar.
We see that for a (110, 96, 384) simulation at $z=0$, 
limited particle density 
becomes a $\gtrsim 2$\% problem only at $k\gtrsim 8\ihmpc$.  At $z=1.5$
it is a more serious problem, surpassing 3\% at $k\simeq 4\ihmpc$ and
quickly diverging (although note the expanded axis scale -- the error 
is actually only 15\% at $k=10\ihmpc$).

A potential solution to this problem is to subtract, after correcting
for the mass assignment smoothing,
Poisson shot-noise power $P_{\rm noise}(k)=\bar{n}^{-1}$, 
where $\bar{n}=(P/L)^3$ is the mean
particle density \citep{1994MNRAS.270..183B,1995MNRAS.274.1049B}; however, 
as others have noted 
\citep{1995MNRAS.274.1049B,2003MNRAS.341.1311S,2005astro.ph..3106S}, 
the idea that the effect
of finite particle density is to add this white noise component is only a
guess, not something that can be assumed to hold, and in fact it is known
not to hold at early times for a uniform grid start. 
Figure \ref{particlenumber} calls into question the idea 
that subtracting white shot-noise is ever useful for high precision 
calculations. 
(See Figure 11 of \cite{2005astro.ph..3106S} for an enlightening  
plot of the evolution of the particle
discreteness power starting from a fixed grid -- we produced a similar figure, 
but \cite{2005astro.ph..3106S}'s is
similar enough that it is not worth including ours in this paper.)
Note that when we say that subtracting Poisson noise is not useful 
for high precision calculations, this does not mean it never leads to
an improvement in accuracy -- it sometimes does -- the problem
is that there is no clearly identifiable regime where the correction is
both significant and accurate enough for high precision work.  
This should probably be regarded as a well-known fact (e.g.,
\cite{1995MNRAS.274.1049B} conclude that the discreteness corrections
they discuss can not be applied consistently, and simply resort to 
using enough particles to make them negligible), but it is worth 
reiterating. 
It seems unlikely that a glass start \citep{2003MNRAS.341.1311S} will
lead to perfectly stable, non-interacting, discreteness power either.  
Ultimately, direct tests of the convergence of observable statistics with 
increasing particle density for different 
methods of setting up the initial conditions 
\citep{2003MNRAS.341.1311S,2004astro.ph.11607J,2005astro.ph..3106S}, 
and possibly different methods for correcting for discreteness, are the only 
way to determine which method works best.  

Figure \ref{particlenumber}(b) shows the same comparison as 
Figure \ref{particlenumber}(b), reduced in scale by a factor of two, i.e.,
(55, 96, 384) is compared to (55, 192, 384), to estimate the accuracy
of a (110, 192, 768) simulation (there is some noise in this comparison
so we have averaged two realizations of each size simulation).
The results are much better, with accuracy
better than 2\% at $z=0$, and better than 4\% for $z=1.5$ (better
than 2\% for $k\lesssim 6\ihmpc$).  Note that the
apparent helpfulness of Poisson noise subtraction at $z=1.5$ is probably
coincidental, as it quickly becomes an over-correction at
$k>10\ihmpc$.

Our default in this paper is to {\it not} subtract shot noise.

\subsection{Force resolution}

\begin{figure}
\centerline{\psfig{file=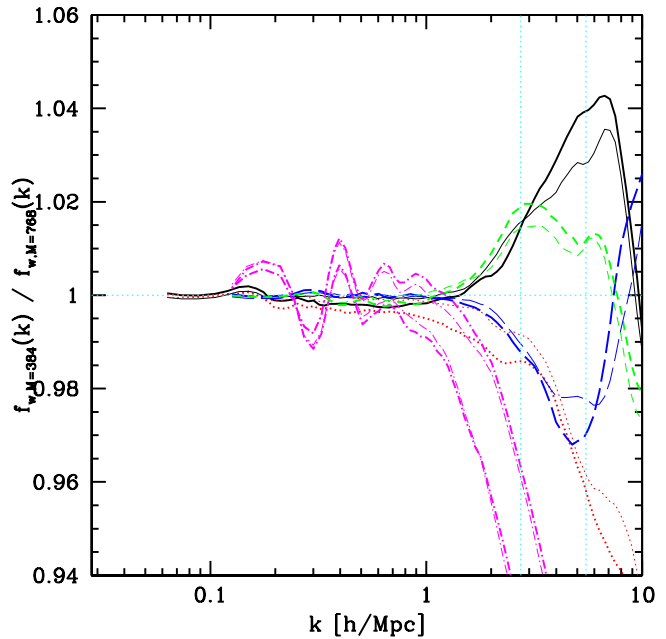,width=3.5in}}
\caption{
Change in the effect of $w$, $f_{w=-0.5}(k)\equiv P_{w=-0.5}(k)/P_{w=-1}$, 
with force resolution. 
Black/solid (red/dotted) lines are power at 
$z=0$ ($z=1.5$) in $L=110\hmpc$ simulations with $(P=96)^3$ particles
(thick lines) or $P=192$ (thin lines). 
The denominator is for mesh $M=768$, with numerator $M=384$.
The blue/long-dashed (green/short-dashed) lines show the same comparisons
for $L=55\hmpc$ simulations (the effect appears to change sign, i.e., the
curves are not mislabeled).
For reference,
magenta/dot-dashed lines show the ratio of power spectra [i.e., 
simply $P(k)$, not $f_w(k)$] for $w=-1$, comparing $M=384$ to 
$M=768$, for an $L=55\hmpc$ box with $P=96$ (thick) or $P=192$ (thin)
(the poorer agreement in each case is $z=0$, better is $z=1.5$).
The vertical cyan/dotted lines show $k=(96,192)\pi/110 \hmpc$.
}
\label{forcemesh}
\end{figure}

Figure \ref{forcemesh} shows the effect of increasing the force mesh 
resolution, for a fixed number of particles.
Comparing $L=110\hmpc$ simulations with $(M=384)^3$ force mesh cells to 
$M=768$, we see that the error on the former is as large as 4\%  at $z=0$
and 7\% at $z=1.5$.  The errors fall to $<3$\% for a similar comparison using
$L=55\hmpc$ boxes.  In Figure \ref{forcemesh}, we break from our general 
policy of plotting only ratios of different models to show the lack of
convergence
of the absolute power spectrum, $P_{M=384}(k)/P_{M=768}(k)$.  This shows the
value of computing ratios -- the better than 3\% agreement in $f_{w=-0.5}(k)$
for the two meshes occurs in spite of a suppression of the absolute power
by as much as 36\%.
 
\subsection{Box size}

Insufficient box size can cause two problems:  simple random 
fluctuations around the mean of a statistic due to limited volume, 
i.e., sample variance;
and systematic errors in the mean of a statistic 
due to missing couplings to large-scale
modes (or even small-scale modes missing due to limited $k$-resolution).  
The first of these can be eliminated by averaging over sufficient
realizations of the initial conditions while the second can not 
(although various methods have been proposed to improve the results,
e.g., \cite{2005astro.ph..3106S}). 
\cite{2005MNRAS.tmp..232B} discuss requirements on simulation 
box size, but not
for the power spectrum -- note that the requirements on numerical 
parameters will generally be different for different statistics
(one advantage of the power spectrum is that it is not directly
sensitive to structure on scales larger than the box).

\begin{figure}
\centerline{\psfig{file=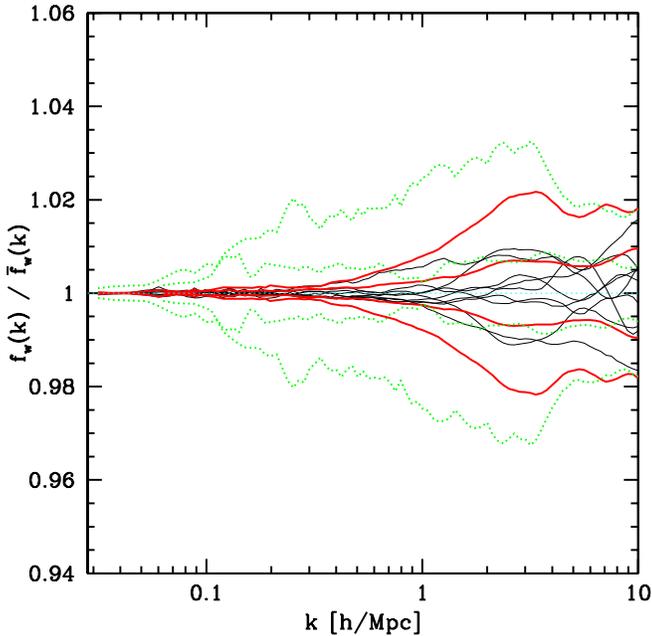,width=3.5in}}
\caption{
Level of random fluctuations in the fractional effect of $w$ on the
non-linear power.   
Black, thin lines show $f_{w=-0.5}(k)$ from $L=220\hmpc$ simulations 
with nine different random seeds, each
divided by the average of $f_{w=-0.5}(k)$ over all nine, at $z=0$.
The inner thick red/solid (green/dotted) lines show bin-by-bin the standard 
deviation of 
the nine around their mean (i.e., the error when using a single realization) 
at $z=0$ ($z=1.5$).  The outer thick lines are the same but for 
$L=110\hmpc$ simulations.
}
\label{random}
\end{figure}

Figure \ref{random} shows that a single $L=220~\hmpc$ simulation is 
sufficient to 
compute the fractional difference in power between $w=-0.5$ and $w=-1$ to
better than 1\% rms statistical error.  A single 
$L=110~\hmpc$ simulation would achieve about 2\% precision at $z=0$ and
3\% at $z=1.5$.
In principle this Figure can be sensitive to the binning in $k$ 
but in practice it is not because the
errors in nearby bins are strongly correlated. 

\begin{figure}
\centerline{\psfig{file=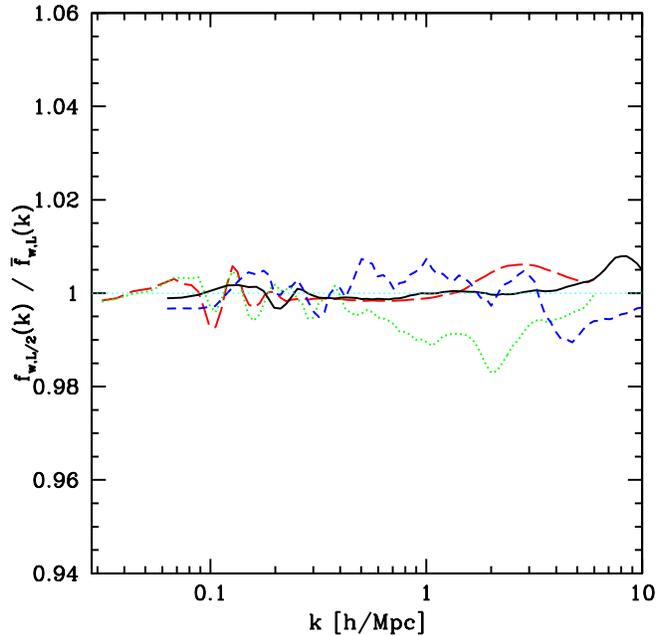,width=3.5in}}
\caption{
Effect of box size on the fractional effect of $w$ on the non-linear power.   
The black/solid (blue/short-dashed) line shows at $z=0$ ($z=1.5$) the ratio of
$f_{w=-0.5}(k)$ computed by averaging over nine $L=110\hmpc$ simulations 
to the result from $L=220\hmpc$ simulations similarly averaged.
The red/long-dashed (green/dotted) line shows at $z=0$ ($z=1.5$) similar
ratios for $L=220\hmpc$ compared to $L=440\hmpc$.  Note that different
sized simulations do not have matching grids in $k$, so these comparisons
involve some interpolation.
}
\label{boxsize}
\end{figure}

Figure \ref{boxsize} addresses the systematic error possibility 
by comparing $L=440\hmpc$, $220\hmpc$, and $110\hmpc$ boxes, with
the power in each case averaged over 9 realizations with different
seeds.
Remarkably, there is no significant sign of systematic error, even 
in the $L=110\hmpc$ calculation (the error in the absolute power is 
larger).  The maximum disagreement
in Figure \ref{boxsize}, $<2$\% between the $L=440\hmpc$ and $L=220\hmpc$
simulations at $z=1.5$, appears not to be strictly a boxsize effect at all,
but rather a coupling between boxsize and limited particle density 
(the particle noise appears to whiten more quickly in the larger box).
For these comparisons we have matched the particle density and force 
resolution between the two box sizes, i.e., we compare (440, 192, 768)
to (220, 96, 384), and (220, 192, 768) to (110, 96, 384). 

\subsection{Time steps}

\begin{figure}
\centerline{\psfig{file=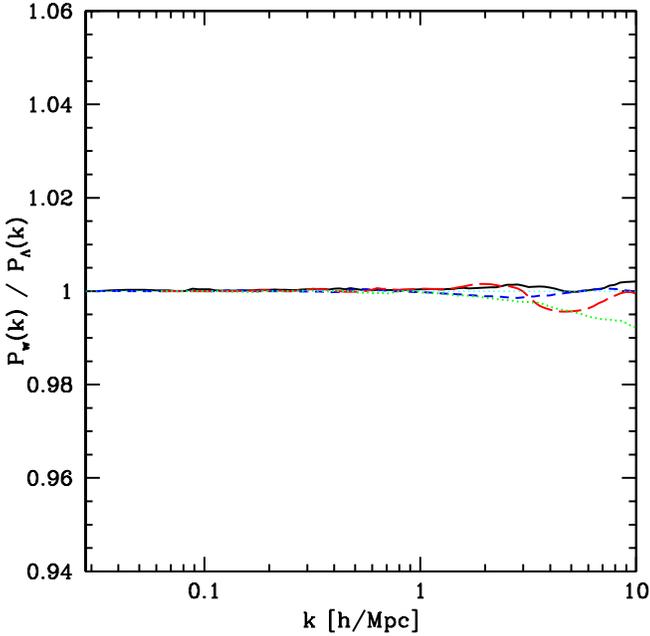,width=3.5in}}
\caption{
Effect of time step size on the fractional effect of $w$ on the 
non-linear power.   
The black/solid (blue/short-dashed) line shows at $z=0$ ($z=1.5$) the ratio of
$f_{w=-0.5}(k)$ computed from (220, 192, 768) 
simulations with $\sim 80$ time steps 
($\sim 30$ for $0.0<z<1.5$) to $\sim 210$ ($\sim 120$) time steps.   
The red/long-dashed (green/dotted) line shows the same comparison for
(110, 192, 768) simulations.
}
\label{timestep}
\end{figure}

For our main grid we used $\sim 210$ time steps to evolve the simulations
($\sim 120$ for $0.0<z<1.5$).  
Reducing this to $\sim 80$ 
($\sim 30$) leads to the $<1$\% change shown in Figure \ref{timestep}.  
It appears that we could accelerate our grid calculation by a factor of a 
few by relaxing our standard time step restrictions.

\subsection{Power spectrum computation}

\begin{figure}
\centerline{\psfig{file=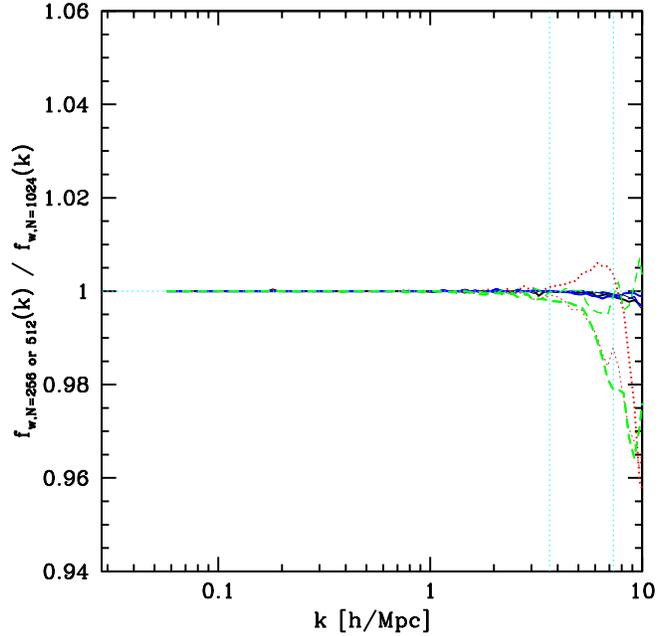,width=3.5in}}
\caption{
Change in effect of $w$, $f_{w=-0.5}(k)\equiv P_{w=-0.5}(k)/P_{w=-1}$, when
we change the resolution of the grid that we use to compute the power 
spectrum for $L=110\hmpc$ simulations.
The denominator is always power computed using $(N=1024)^3$ cells.
For $P=192$, $M=768$, black (solid) show the
change for $N=512$ and red (dotted) $N=256$, while
for $P=96$, $M=384$ we use blue (long-dashed) and green (short-dashed)
to show the same two $N$'s, respectively.
The thick lines show $z=0$, thin lines show $z=1.5$. 
The vertical cyan (dotted) lines mark $(128, 256)\pi/L$. 
}
\label{wgridtest}
\end{figure}

We generally use an $(N=1024)^3$ grid for the power spectrum computation, with
a simple correction for the CIC smoothing, which we see in Figure 
\ref{wgridtest} is essentially
exact (better than 0.5\% as tested using measurements with different $N$) 
out to $\sim 0.7~k_{\rm Nyq}$, where $k_{\rm Nyq}\equiv \pi N/L$ 
(see \cite{2004astro.ph..9240J} for a discussion of aliasing).
Note that if we are looking at a simple ratio of raw power in two simulations
of the same size, the CIC smoothing correction we apply has no effect, because
it is multiplicative
(assuming we are not subtracting shot-noise).
Figure \ref{wgridtest} shows that a grid spacing
$\Delta x\sim 0.2~\hmpc$
is sufficient to introduce essentially no error into 
our computation at $k<10~\ihmpc$,
independent of the particle density.

\section{Simulation grid and fitting formula}

Our limited quantitative goal in this paper is to provide a module that 
can be grafted onto \cite{2003MNRAS.341.1311S} to account for varying $w$.  

We found in \S 3 that to reduce finite particle density and
force resolution errors to 2-3\%, we need resolution equivalent to 
(110, 192, 768).  An $L=110\hmpc$ box is sufficient to achieve percent
level systematic accuracy, and $\sim 3$\% statistical precision, i.e.,
(110, 192, 768) simulations are sufficient to essentially solve the problem
of computing the effect of $w$ to a few percent, especially if we average
over a few different realizations of the initial conditions, and considering
that the maximum errors are generally near $10\ihmpc$, where baryons 
probably limit our precision anyway.   The code we describe in this section
is based on a grid of
(110, 192, 768) simulations, averaged over four realizations of the
initial conditions for each grid point.
At this point it would be straightforward to perform a grid
of larger simulations to meet the numerical requirements more comfortably,
but given the generally limited scope of this paper, we have deferred 
this to future work.

Our grid of models is motivated by the idea of Taylor expanding the dependence
of $\weffect$ on the other parameters around a central model.  
The central model and positive and negative variations are
$\sigma_8=0.897\pm 0.097$, $\Omega_m=0.281\pm 0.070$, 
$\Omega_b=0.0462 \pm 0.0052$, $h=0.710\pm 0.066$, and 
$n=0.980\pm 0.065$ 
(motivated by the best fit
and $3 \sigma$ errors from \cite{2005PhRvD..71j3515S})  
To be clear:  we are only 
varying these parameters individually, not in combinations, i.e., the 
grid does not have $3^N$ points.  For the two most important parameters,
$\Omega_m$ and $\sigma_8$, we add the four possible joint variations to 
the grid. 
For each variation of the non-$w$ parameters, 
we run models with $w=$-0.5, -0.75, -1.0, and -1.5.
We extract the power spectrum at $z=$1.5, 1.0, 0.5, 0.25, and 0.0.
We do not advocate this kind of grid for a more general simulation 
project, where a set of simulations guided by CMB constraints should
be most efficient \citep{2004APh....22...19W}.

It would probably be straightforward to extend \cite{2003MNRAS.341.1311S} by 
modifying their fitting 
functions $f_i(\Omega_m)$ to depend on $w$, but we take the less
sophisticated approach of describing the change in power relative 
to $w=-1$ by a 
multi-polynomial function of the cosmological parameters.
If $\mathbf{p}$ is the vector of cosmological parameters, which
we take to include $z$, the formula for the correction factor is
\begin{eqnarray}
\ln\left[\frac{P_{\rm NL}(w)/D^2(w)}
{P_{\rm NL}(w=-1) /D^2(w=-1)}\right](k,\mathbf{p})
&=&\\ \nonumber
\left(\prod_{i=1}^{N_p}\sum_{\nu_i=0}^{N_i}p_{i}^{\nu_i}\right)
A_{\nu_1\nu_2\nu_3...\nu_{N_p}}(k)&,&
\label{interpeq}
\end{eqnarray}
where $N_p$ is the number of cosmological parameters, $N_i$ is the order of
polynomial to use for each of them, and
$A_{\nu_1\nu_2\nu_3...\nu_{N_p}}(k)$ are coefficients to be determined
by a least-squares fit to simulations.  $D(w)$ is the linear growth factor, 
normalized to 1 at $z=0$ (we divide out the growth factor in 
Equation \ref{interpeq} to remove the relatively trivial linear
evolution with redshift). 
Note that this formula includes many cross-terms
for which we do not have simulations in our grid -- their coefficients
are set to zero when the fit is performed using singular value decomposition
(we use Equation \ref{interpeq} because it is easy to write down and implement
in code, and extends automatically when additional simulations become 
available).  
We measure $P_{\rm NL}(k)$ from the simulations in 
bands spaced by $\Delta \log_{10} k = 0.1$, and determine a separate set 
of $A$s for each band.
Once the $A$s are determined, $P_{\rm NL}$ for any model is computed by 
simply plugging the desired parameters into Equation \ref{interpeq} and   
multiplying the result by the non-linear power in
the corresponding $w=-1$ model, and the appropriate linear growth
factors.
Our code quantifying the results, with an example showing how to use it, 
can be found at
http://www.cita.utoronto.ca/$\sim$pmcdonal/code.html, under the name
``wcorrector.'' This code is only tested for $k<10 \ihmpc$.  The user
who needs to integrate to higher $k$ should 
absorb the uncertainty in the extrapolation into the uncertainty they 
are assuming from baryon effects. 

Note that we have not 
separated the dependence of the non-linear power on $\Omega_m$
through its effect on past non-linear growth from its effect 
through 
the transfer function (except in Figure \ref{omegasmith}).  
Separating these would probably be useful 
in the future, so that formulas can be used with 
arbitrary linear power spectrum without fear that spurious effects
of $\Omega_m$ will be generated.  
However, because we have used fully accurate transfer
functions from ``lingers'' \citep{2001ApJS..137....1B}, the
coupling between the two influences of $\Omega_m$ 
can have no effect on calculations
within the standard $\Lambda$CDM model.  Additionally, we showed
that the effect of changing the power spectrum shape on $f_w$ 
at fixed $\sigma_8$ is negligible, so it can probably be safely
assumed that the effect of $\Omega_m$ that we see is coming
through the past growth.

The reader may ask whether a more thoughtful, theoretically motivated fitting
formula could be more general and efficient.  This is possible.  
On the other hand, a
drawback of such formulas is that they trap the user into a limited functional
form for the various dependences.  When a carefully constructed fitting 
formula is found not to fit to the required level of precision, it is not 
generally straightforward to extend it.  It is simple to extend 
Equation \ref{interpeq} to fit any set of simulations.  Furthermore, as
we saw in the case of the 
\cite{1999ApJ...521L...1M} formula, it is
easy in these cases to be deceived into thinking that they are more 
generally applicable than they really are, i.e., to think of them as having
genuine predictive power rather than simply being a method of interpolation
between simulation results.  Equation \ref{interpeq} makes the interpolatory
nature of these fitting formulas explicit.  
Ultimately, a hybrid approach may be optimal.  A physically motivated
fitting formula could be used to remove much of the gross parameter
dependence, with a general formula like Equation \ref{interpeq} used
to make corrections for the imperfections in the fitting formula.
The hope would be that this would allow a sparser calibration grid
than would otherwise be needed.   

\section{Conclusions}

We have isolated the effect of changing $w$ on the non-linear power
spectrum of mass density fluctuations by comparing simulations
with identical linear theory density fields at the observed redshift.      
We focused on this definition of the effect of $w$ (i.e., fixed linear
theory power and other parameters at $z=0$) simply because it has not
been carefully considered in the past, and this complements predictive
formulas calibrated only for $w=-1$ (e.g., \cite{2003MNRAS.341.1311S}).
The change in power relative to $w=-1$ is $\lesssim 10$\% for $k<1\ihmpc$
(for $-0.5<w<-1.5$), but rises to
12\% by $k=10 \ihmpc$ in a model with $w=-0.75$, 
and $\sim 33$\% for $w=-0.5$ (at $z=0$).  Among the
other cosmological parameters, the size
of the effect is primarily sensitive to the dark energy fraction,
i.e., $\Omega_m$ in flat models.
The power spectrum normalization (i.e., $\sigma_8$) also has a 
small effect, but the slope/shape of the power spectrum are
irrelevant (as represented by varying $n$, $\Omega_b$, and $h$
at fixed $\sigma_8$).  

Figure \ref{omegasmith} confirms the accuracy of
the formula of \cite{2003MNRAS.341.1311S} for the dependence on $\Omega_m$
in $\Lambda$CDM models, at least once it has been modified to use high  
accuracy transfer functions.

We provide a simple code 
(http://www.cita.utoronto.ca/$\sim$pmcdonal/code.html) 
quantifying the effect of $w$ as 
a function of $k$, $z$, $w$, $\Omega_m$, $\sigma_8$, $n$, $h$, and
$\Omega_b$, to be used as a correction to $P_{\rm NL}(k)$
calculations accurate at $w=-1$.
The dependence of the power spectrum on $w$ should be accurate to 
a few percent for $k\lesssim 10~\ihmpc$.
Our quantitative results may be useful for making more realistic
projections of the future potential to measure $w$ by methods 
sensitive to the non-linear power (primarily weak lensing).
Our code will be appropriate for forecasts of parameter measurements
from future large data sets,  
or parameter determinations using data sets that at least 
include WMAP.
It should be used cautiously for fits to limited amounts of weak 
lensing data alone, since we do not cover extreme parameter values 
(however, by construction the code will only be unreliable in regions
of parameter space that are ruled out by other observations). 

Our ambitions have been quite limited in this paper.  We have not
tried to determine the absolute power in the central model because
that is {\it much} more difficult to simulate precisely than the 
fractional changes we studied here, requiring both larger box sizes
and higher resolution.  Much of the error caused by limitations in 
the simulations cancels when we take ratios of power spectra, a
fact that should make future construction of high precision grids 
of simulation predictions easier.

\bibliography{apjmnemonic,cosmo,cosmo_preprints}
\bibliographystyle{mnras} 

\end{document}